# A Machine Learning Approach for Hierarchical Classification of Software Requirements

Manal Binkhonain[a], Liping Zhao*[b]

[a] *College of Computer and Information Sciences, King Saud University, Riyadh, Saudi Arabia*
[b] *Department of Computer Science, University of Manchester, Manchester, UK*

**Abstract**

**Context:** Classification of software requirements into different categories is a critically important task in requirements engineering (RE). Developing machine learning (ML) approaches for requirements classification has attracted great interest in the RE community since the 2000s. **Objective:** This paper aims to address two related problems that have been challenging real-world applications of ML approaches: the problems of class imbalance and high dimensionality with low sample size data (HDLSS). These problems can greatly degrade the classification performance of ML methods. **Method:** The paper proposes **HC4RC**, a novel ML approach for multiclass classification of requirements. HC4RC solves the aforementioned problems through semantic-role based feature selection, dataset decomposition and hierarchical classification. We experimentally compare the effectiveness of HC4RC with three closely related approaches - two of which are based on a traditional statistical classification model whereas one using an advanced deep learning model. **Results:** Our experiment shows: 1) The class imbalance and HDLSS problems present a challenge to both traditional and advanced ML approaches. 2) The HC4RC approach is simple to use and can effectively address the class imbalance and HDLSS problems compared to similar approaches. **Conclusion:** This paper makes an important practical contribution to addressing the class imbalance and HDLSS problems in multiclass classification of software

*Correspondence to Department of Computer Science, University of Manchester, Manchester, M13 9PL, UK.

*Email address:* mbinkhonain@KSU.EDU.SA (M. Binkhonin), liping.zhao@manchester.ac.uk. (L. Zhao).

requirements.

---

## 1. Introduction

In recent years, machine learning (ML) approaches have achieved visible successes in a wide range of real-world applications, from fake news detection (Agarwal et al., 2020), to opinion mining (Jin et al., 2009), sentiment analysis (Ravi and Ravi, 2015), spam email filtering (Jin et al., 2009), traffic predication (Sarker, 2021), and medical diagnosis (Sidey-Gibbons and Sidey-Gibbons, 2019), to name just a few. Away from these general applications, ML approaches have also attracted great interest in the requirements engineering (RE) community, with more and more RE researchers actively seeking to develop practical ML applications for requirements analysis tasks. Such tasks include requirements classification (Cleland-Huang et al., 2006), requirements prioritization (Perini et al., 2012), requirements detection (Abualhaija et al., 2019), and requirements traceability (Cleland-Huang et al., 2007a). In this paper, we focus on **requirements classification**, a task central and critical to successful software development projects (Glinz, 2007; Chung and do Prado Leite, 2009; Broy, 2015).

A requirement for a software system development project is a statement of what the system should do or how well the system should perform. Examples of requirements are: "The system must provide an online help function" and "It must be possible to completely restore a running configuration when the system crashes". An average software project normally has a few hundreds of requirements (Eckhardt et al., 2016). The complete set of requirements for a specific system is called a "requirements document" or a "requirements specification". Requirements classification is the task of the assignment of a given set of requirements in a document to different categories or classes according to a specific classification scheme. This typically involves classifying each requirement as either a functional requirement (FR) or a non-functional requirement (NFR) (Chung and do Prado Leite, 2009). A NFR can be further classified as a security, reliability, performance, or usability requirement. In the aforementioned examples, "The system must provide an online help function" should be classified as a FR, whereas "It must be possible to completely restore a running configuration when the system crashes" should be classified as a reliability requirement, a specific NFR.

As shown above, requirements are stated in natural language (Zhao et al., 2021), which means that *requirements documents are text documents*. Consequently, ML approaches to text classification (Sebastiani, 2002; Kowsari et al., 2019) can be adapted to requirements classification, whereby we train a requirements classifier with a set of labelled requirements examples (i.e., the training set) (Binkhonain and Zhao, 2019). In other words, *ML approaches to requirements classification are based on supervised text classification*. Furthermore, requirements classification is typically a **multiclass classification** task as it deals with more than two classes (usually more than 10 different classes).

Since the publication of the landmark work by Cleland-Huang et al. (2006) on ML-based requirements classification more than a decade ago, RE researchers have proposed a large number of ML approaches. While most of these approaches are based on traditional ML algorithms (Cleland-Huang et al., 2007c; Ko et al., 2007; Casamayor et al., 2010; Kurtanović and Maalej, 2017; Dalpiaz et al., 2019; Abualhaija et al., 2020; Dias Canedo and Cordeiro Mendes, 2020), more recent proposals are exploring the use of deep learning (DL) models for requirements classification (Hey et al., 2020a; Mekala et al., 2021).

However, regardless of what ML approach is used, a common problem with requirements classification is **class imbalance** in the training data (He and Garcia, 2009), as requirements categories are naturally uneven, usually with a small percentage of categories containing a large percentage of the requirements (Kurtanović and Maalej, 2017; Eckhardt et al., 2016). Class imbalance in requirements classification is known as *relative class imbalance*, as the minority classes are not necessarily rare in their own right but rather relative to the majority classes (He and Garcia, 2009). Relative class imbalance occurs frequently in real-world applications, such as the detection of oil spills in satellite radar images, the detection of fraudulent telephone calls, information retrieval and filtering, and diagnoses of rare medical conditions (Japkowicz and Stephen, 2002).

Imbalanced classes in the training set can cause *imbalanced learning* (He and Garcia, 2009), as ML classifiers will have more examples to learn in the majority classes than in the minority classes (He and Garcia, 2009; Seiffert et al., 2014; Li et al., 2020; Jiang et al., 2013). Consequently, imbalanced classes can lead to misclassification.

There exist different techniques for dealing with imbalanced classes (Li et al., 2020). These techniques generally attempt to reduce the severity of imbalance within the training data by providing a more equivalent statistical representation of the majority and minority classes (Mills et al., 2018). Among them are data sampling (or resampling) techniques (He and Garcia, 2009), used to either remove excessive samples from the majority classes (known as *under-sampling*) or to add more samples to the minority classes (known as *oversampling*). Both over and under sampling techniques have also been used in ML approaches for requirement classification (Kurtanović and Maalej, 2017; Hey et al., 2020a).

However, sampling techniques have their own drawbacks. In particular, oversampling can cause a classifier to over-ft to the minority classes, whereas under-sampling can affect the performance of the classifier on the majority classes, due to the risk of removing good representative samples from these classes (Wang and Yao, 2012; Li et al., 2020). Additionally, oversampling can be an issue for requirements classification, due to the lack of labelled requirements (Alhoshan et al., 2022).

The class imbalance problem can become even worse when combined with the problem of *high dimensionality* and *low sample size* (**HDLSS**) datasets (He and Garcia, 2009; Shen et al., 2022). The HDLSS problem is concerned with the scenario where the sample size $n$ in a training dataset is dramatically smaller than the feature dimension $d$, where ($n << d$) (Shen et al., 2022). HDLSS data can seriously degrade the classification performance of classical statistical methods (Shen et al., 2022).

HDLSS data are common in many real-world applications such as data mining, image processing and computer vision, bioinformatics, and gene expression (Shen et al., 2022). The problem is also present in the training data of requirements classification. The combination of class imbalance and HDLSS data presents a critical challenge to many ML approaches, as HDLSS can amplify the imbalanced data and makes the classifier even more closely or exactly fitted to a specific training set (He and Garcia, 2009; Liu et al., 2017).

To address the HDLSS problem, researchers have turned to feature selection techniques for answers (Wasikowski and Chen, 2009; Sima and Dougherty, 2006; Liu et al., 2017; Huang et al., 2017), as these techniques can not only successfully reduce the dimensions in texts, but also reduce the over-fitting problem (Wasikowski and Chen, 2009; Zheng et al., 2004; Chen et al., 2009;

Yin et al., 2013; Liu et al., 2017; Huang et al., 2017; Fu et al., 2020).

In this paper, we propose a novel ML approach for requirements classification. The proposed approach, called **HC4RC** (Hierarchical Classification for Requirements Classification), aims to address the class imbalance and HDLSS problems by means of three novel techniques, namely *Semantic Role-Based Feature Selection* (**SR4FS**), *Dataset Decomposition* and *Hierarchical Classification.* Specifically, SR4FS addresses the HDLSS problem in the requirements training data using a small set of *semantic roles*, such as agent, action and goal (Gildea and Jurafsky, 2002), so as to reduce the number of word features on each requirement statement. Dataset Decomposition and Hierarchical Classification work together to handle the class imbalance problem in requirements classes, with the former for rebalancing the training set by decomposing it into two approximately balanced sets and the latter for performing hierarchical classification on the decomposed datasets.

We continue this paper as follows: Section 2 justifies the novelty of the above three techniques by reviewing related techniques used in general text classification applications as well as in requirements classification. Section 3 explains the principles of each of these techniques in detail, how we combine them into a coherent approach, HC4RC. Section 4 describes the procedures and methods by which we experimentally compare HC4RC with three closely related ML approaches, whereas Section 5 presents and analyzes the experiment results. Our implementation code for all the compared approaches used in our experiments is provided at the Code Ocean platform (`https://codeocean.com`, with `doi:10.24433/CO.6887783.v1`). Then Section 6 discusses the validity and limitations of the proposed approach and our evaluation methods. Finally, Section 7 concludes our paper and summarises our contributions.

## 2. Related Work and Our Contributions

In this section, we review some prominent techniques that have been used to solve the class imbalance and HDLSS problems. In Section 2.1, we review common feature selection techniques used in text classification as solutions to the HDLSS problem; in Section 2.2, we present feature selection techniques used in requirements classification and justify our novel contribution. In Section 2.3, we present common data re-balancing techniques used in classification tasks as solutions to the class imbalance problem; finally, Section 2.4 reviews re-balancing techniques used in requirements classification and justify our novel contribution.

*2.1. Feature Selection for Text Classification*

A major challenge of text classification is high dimensionality of the feature space (Deng et al., 2019). A text document usually contains hundreds or thousands of distinct words that are regarded as features for classifiers, however, many of them may be noisy, less informative, or redundant with respect to class labels. This may mislead the classifiers and degrade their performance in general (Sebastiani, 2002; Deng et al., 2019). Therefore, feature selection must be applied to eliminate irrelevant features, so as to reduce the feature space to a manageable level, thus improving the efficiency and accuracy of the classifiers used (Kowsari et al., 2019; Deng et al., 2019). In this paper, feature selection plays a specific role at addressing the HDLSS problem (Wasikowski and Chen, 2009; Sima and Dougherty, 2006; Liu et al., 2017; Huang et al., 2017).

Feature selection techniques for text classification broadly fall into three categories: *syntactic word representation*, *weighted words* and *semantic word representation* (Kowsari et al., 2019). The most basic form of syntactic word representation feature selection is *n-gram* (e.g., 1-gram, 2-gram, 3-gram, etc.), which is a set of n-words occurring consecutively in a text. Other syntactic word representations include syntactic features on the text, such as part-of-speech (POS) tags (Deng et al., 2019).

The most common weighted word feature selection techniques are *TF* (Term Frequency), *TF-IDF* (Term Frequency-Inverse Document Frequency) and *BOW* (Bag-of-Words). These techniques use word frequency to calculate the weight (importance) of each word in a text (Kowsari et al., 2019).

The current approach for semantic word representation for feature selection is *word embeddings*, where each word or phrase from the vocabulary is mapped to a $N$ dimension vector of real numbers (Kowsari et al., 2019). Examples of common word embedding techniques are Word2Vec (Mikolov et al., 2013a,b), GloVe (Pennington et al., 2014) and FastText (Bojanowski et al., 2017).

However, each of these three types of feature selection technique has its own limitations. For example, n-gram relies on an extensive dictionary of words to identify features, whereas word frequency techniques such as BOW will fail if none of the words in the training set are included in the testing set. Word embeddings require a large corpus to train an embedding model.

*2.2. Feature Selection in Requirements Classification and Our Contribution*

Due to the domain-specificity nature of the requirements texts and lack of labelled data (Ferrari et al., 2017), requirements classification normally employs *ad hoc* techniques for feature selection. These include *keyword-based* (Cleland-Huang et al., 2007c), which uses a dictionary of requirements keywords for feature selection; *syntactic feature-based*, which derives word features from various syntactic features, such as POS tags, n-grams, verbs, and syntactic dependency rules (Kurtanović and Maalej, 2017; Abualhaija et al., 2020; Dalpiaz et al., 2019). However, as these feature selection techniques also involve the frequency analysis of features, they suffer similar drawbacks as the aforementioned techniques.

In this paper, we propose a simple semantic word representation technique for feature selection. The technique, **SR4FS**, uses a small set of semantic roles to identify meaningful and representative word features from requirements statements. Semantic roles, also known as thematic roles, are the various roles or positions that words in a sentence may play with respect to the action or state described by a governing verb, commonly the sentence's main verb (Gildea and Jurafsky, 2002). The set of semantic roles formulated by us is based on our knowledge of the semantic concepts of requirements (Letsholo et al., 2013), rather than the frequencies of the words in the dataset. Consequently, SR4FS is independent of the frequency of words in the training set and the size of the training set. This novel feature selection technique is presented in Section 3.

*2.3. Class Rebalancing and Hierarchical Classification*

Most existing techniques to the imbalanced learning problem are designed to address binary-class problems, in which imbalances exist between two classes (He and Garcia, 2009), but these solutions are found to be less effective or even cause negative effects on multiclass classification tasks Wang and Yao (2012). Existing solutions for multiclass imbalance problems are very limited, among which are the aforementioned *data sampling* (oversampling, under-sampling and a combination of both) and *data decomposition* techniques (Feng et al., 2018; Li et al., 2020).

Data sampling techniques either increase (oversampling) or decrease (under-sampling) the number of instances in the sampled classes. Data sampling can be carried out randomly - *random sampling* - or with targeted majority or minority classes. However, while oversampling increases the risk of over-fitting to the minority classes, under-sampling is sensitive to the number of minority classes and can cause performance loss on majority classes Wang and Yao (2012), as it may remove good representative instances from majority classes, which ultimately misleads the classification (Li et al., 2020).

Data decomposition techniques generally entail decomposing a multiclass classification problem into a series of smaller two-class sub-problems (He and Garcia, 2009) and then applying binary-class classification to these sub-problems. These techniques include One-Versus-All (OVA) (also known as One-Versus-Rest), One-Versus-One (OVO) (Li et al., 2020) and a class decomposition technique proposed by Yin et al. (2013). However, data decomposition techniques can aggravate imbalanced class distributions (Żak and Woźniak, 2020; Li et al., 2020) and the combined results from the binary classifiers learned from the different sub-datasets can cause potential classification errors, as each individual classifier is trained without the full knowledge of the entire dataset (Feng et al., 2018).

In recent years, *ensemble-based imbalance learning* techniques have been adapted to multiclass imbalance problems, with positive results. For example, Wang and Yao (2012) show that a boosting-based ensemble that combines AdaBoost with random oversampling can improve the prediction accuracy on the minority class without losing the overall performance compared to other existing class imbalance learning methods. Feng et al. (2018) show that a bagging-based ensemble that combines margin ordering with under-sampling can improve a classifier's recognition of minority class instances without decreasing the accuracy of majority class. However, all these techniques use class decomposition to convert a multiclass imbalance problem into a series of binary-class sub-problems and then apply a set of binary classifiers to classify these sub-problems (Galar et al., 2011).

However, although ensemble-based techniques can improve classification performance on imbalanced classes, their success depends on the creation of diverse classifiers while maintaining their consistency with the training set (Galar et al., 2011) and this is not easy (Brown et al., 2005), as the concept of diversity is still ill-defined in ML (Galar et al., 2011).

Originally designed for classification with hierarchical class structures, *hierarchical classification* (Kiritchenko et al., 2006) has also shown promise for class imbalance problems in text classification (Ghazi et al., 2010; Zheng and Zhao, 2020). In hierarchical classification, classes are organized into a tree structure with levels and nodes (Kiritchenko et al., 2006). Accordingly, the classification task is also divided into a set of sub-tasks corresponding to the number of nodes. The construction of a classification hierarchy can be informed by domain knowledge (e.g., relationships between the classes (Ghazi et al., 2010)) or constraints (e.g., cost-sensitive factor (Zheng and Zhao, 2020)), with an aim to address the class imbalance problem.

*2.4. Class Rebalancing in Requirements Classification and Our Contribution*

In requirements classification, we only found two approaches that have explicitly addressed the class imbalance problem, one by Kurtanović and Maalej (2017) and another by Hey et al. (2020a). These approaches all adopt data sampling techniques, using oversampling for the minority class and under-sampling for the majority class. The sampling techniques used in these approaches are for binary classification, whereby the requirements are classified into functional and non-functional requirements. These approaches have not addressed the class imbalance problem in multiclass classification tasks, which are inherent to requirements classification.

In this paper, we propose **HC4RC**, a novel approach for multiclass imbalanced learning that combines *dataset decomposition* with *hierarchical classification* (Ghazi et al., 2010). Under this approach, we first decompose the training dataset into two balanced subsets, one with the majority classes and another with the minority classes. In doing so, we divide a "flat", imbalanced classification problem into a hierarchy of two smaller, balanced problems. We then train a hierarchy of three classifiers, one binary and two multiclass classifiers and use them to perform three sub-classification tasks. While the binary classifier classifies each requirement into either the majority class set or the minority class set, the two multiclass classifiers each perform classification in its corresponding subset. The basic idea of this technique is to partition the training dataset into two subsets so as to reduce between-class imbalances within each subset, as the classes in each subset are relatively balanced. The decomposition step is similar to solving a two-class (binary) imbalanced problem. This novel approach is presented in Section 3.

## 3. The HC4RC Approach

As stated early, the HC4RC approach uses three novel techniques to solve the class imbalance and HDLSS problems in requirements classification. These techniques are summarised below:

- **Semantic Role-based Feature Selection**. This technique uses a small number of semantic roles to identify most relevant semantic

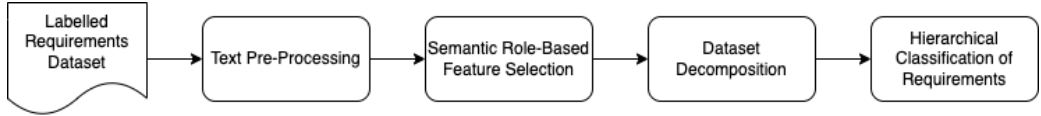

Figure 1: The training process of HC4RC and its key techniqiues.

features from the requirements, to address the high dimensionality and low sample size problems.

- **Dataset Decomposition**. This technique aims to rebalance a given training dataset, by annotating it into two approximately balanced datasets, with one containing the majority classes and another the minority classes.

- **Hierarchical Classification**. This technique works with Data Decomposition, to perform hierarchical classification on the decomposed datasets.

These techniques are organized as a series of steps and integrated into a coherent training process, as shown in Figure 1. These steps and the principles and rationale behind these techniques are described in the sections below.

*3.1. Text Pre-Processing*

This is a necessary first step in text classification, as text data contain many noises and unwanted words. The purpose of this step is to clean and standardize the text so that the text can be processed in the subsequent steps (Sarkar, 2016; Dias Canedo and Cordeiro Mendes, 2020). Various natural language processing (NLP) techniques are available for text pre-processing. Some commonly used NLP techniques for requirements classification are described in a survey by Binkhonain and Zhao (2019).

In HC4RC, we apply the following NLP techniques to the requirements text, in the order of, tokenization, lowercase conversion, lemmatization, and stop words and short words removal (short words are the words containing fewer than three characters).

*3.2. Semantic Role-Based Feature Selection*

This step aims to select a small number of most relevant features for each requirement statement. To do so, we apply our semantic-role based feature selection technique, **SR4FS**. Below we introduce the set of semantic roles used in our approach and the principles behind these roles. We then explain how we can identify them from requirements statements.

A *semantic role* is a word or phrase in a sentence that plays a certain role in relation to the sentence's main verb. There are many kinds of semantic role (Gildea and Jurafsky, 2002), but we only adopt six of them for our SR4FS, as they are similar to the concepts used in requirements modelling (Rolland and Proix, 1992). These six semantic roles are introduced here:

1. **Agent** - the volitional causer of an event or action (Jurafsky and Martin, 2020). This role is played by the main *subject* in a sentence For example, in the requirement statement: "The system shall send a verification email to the user when they log on to their account from an unfamiliar computer", the word "system" is an agent. An agent is also called an **actor** (Rolland and Proix, 1992).
2. **Action** - the cause of an action, event or state. This role is fulfilled by the *verb* of a sentence. For example, in the requirement statement: "The system shall send a verification email to the users when they log on to their account from an unfamiliar computer", the words "send" and "log on" play the action role.
3. **Theme** - the participant most directly affected by an event or action (Jurafsky and Martin, 2020). This role is played by the *direct object* in a sentence. For example, in the requirement "The system shall send a verification message to the users when they log on to their account from an unfamiliar computer", the word "message" takes the theme role. In requirements modelling, the theme role is also referred to as the **key object** (Sutcliffe and Maiden, 1998).
4. **Goal** - the destination of an object of a transfer event or an action (Jurafsky and Martin, 2020). This role is fulfilled by the *indirect object* in a sentence. For example, in the requirement: "The system shall send a verification email to the users when they log on to their account from an unfamiliar computer", the word "user" is the goal. In requirements modelling, the goal describes a future, required state which the system should satisfy, maintain or sometimes avoid (Sutcliffe and Maiden, 1998).
5. **Manner** - the manner in which an action is taking place (Xue, 2008). This role is fulfilled by an *adjective*, *adverb*, *determiner*, or *preposition phrase*. For example, in the requirement "The system should be easy to use", the adjective phrase "easy to use" plays the manner role.

Table 1: Six Semantic Roles of SR4FS and their Mapping to Corresponding Grammatical Features.

| Semantic Roles | Grammatical Features | Mapping Rules |
| --- | --- | --- |
| **Agent** | 1. Subject | If a term is the subject of the head verb, it corresponds to an **agent**. |
| **Action** | 2. Action Verb | If a term is the verb and its head is verb, it corresponds to an **action**. |
| **Theme** | 3. Direct Object | If a term is the direct object of the main verb, it corresponds to a **theme**. |
| **Goal** | 4. Indirect Object | If a term is an indirect object of a dative preposition, it corresponds to a **goal**. |
| **Manner** | 5. Adverb; 6. Adjective; 7. Determiner; 8. Proposition Phrase | If a term is an adjective, adverb, or determiner, this term and its headwords represent a **manner**; else, if a term is a preposition (e.g., from, with, without, after), then the preposition and all its dependents correspond to a **manner**. |
| **Measure** | 9. Adverb; 10. Number or Quantity | If a term is a named entity (e.g., data, time, percent, money, and cardinal), then the term and all its dependents represent a **measure**; else, if the term is an adverb, this term and its headwords are mapped onto a **measure**. |

6. **Measure** - the degree of control by the action or the quantification of an event (Jurafsky and Martin, 2020). This role is typically fulfilled by an *adverb* (e.g., rather), a *number* or a *quantity* (e.g., 99%). For example, in the requirement, "The system must be available to the users 98% of the time every month during business hours", the percentage "98%" plays the role of measure.

The above semantic roles are sufficient to answer a range of questions in requirements analysis, as they cover the concerns of: "*Who (**agent**) did (**action**) what (**theme**) to whom (**goal**), how (**manner**) and how much (**measure**)*". The underlying words of these roles can thus serve as relevant features to requirements classification. In particular, subjects, verbs and objects are highly relevant to the identification of FRs, whereas adjectives, adverbs and quantities are relevant to NFRs.

As can be seen, semantic roles can be mapped onto different parts of speech and grammatical features in sentences. Consequently, they can be automatically identified using NLP tools such as POS tagging, dependency parsing and named entity recognition (NER). The mapping rules between the aforementioned six semantic roles and their corresponding POS tags and grammatical features are presented in Table 1.

SR4FS automatically performs feature selection in two steps:

1. Processes each requirement statement in the training set using a POS tagger, dependency parser and NER tagger.
2. Extracts the POS tags and grammatical features from the above process, and maps them onto semantic roles using the mapping rules.

The features selected from SR4FS will then be manually checked to correct any errors or inaccuracies from the automatic process. The manual checking is needed due to: 1) NLP tools have yet to achieve 100% accuracy[1]; 2) NER tools perform worse on recognising domain-specific entities; and 3) the relationship between a semantic role and its underlying syntactic realisation is not a strict one-to-one mapping.

*3.3. Dataset Decomposition*

The process of Dataset Decomposition aims to decompose a *flat*, imbalanced training set into two subsets with balanced numbers of requirements. However, instead of physically splitting the training set, the process assigns a label to each requirement in the training set to indicate if the requirement belongs to the majority or minority subset. This process consists of these steps:

1. Sort the classes in the training set in descending order, based on the number of requirements in each class.
2. Starting from the top of the list, for each class, assign each requirement in the class a "$maj$" label, to denote that the requirement belongs to the majority class subset. This labelling process ends when the number of requirements in the majority class subset is at least half of the total number of requirements in the training dataset.
3. Finally, assign each remaining requirement a "$min$" label, to denote that the requirement belongs to the minority class subset.

This decomposition process thus divides the original *flat* classification task with an imbalanced dataset into two balanced subtasks, which can then be solved by a hierarchical classification approach, described below.

---

[1]The current state-of-art POS tagger (e.g., *spaCy*) can only achieve a 95.1% accurate whereas the current state-of-art NER (*spaCy* and *RoBERTa*) can achieve an 89.8% accurate (https://spacy.io/usage/facts-figures).

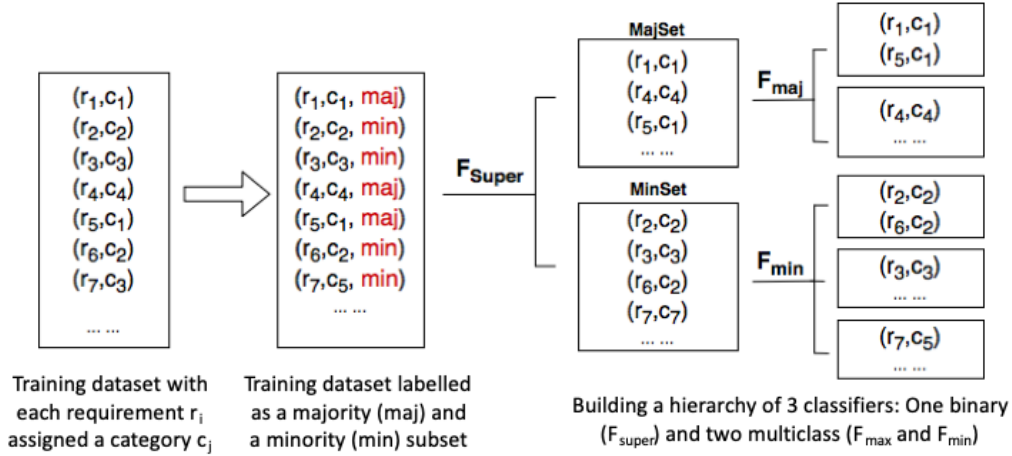

Figure 2: Dataset Decomposition and Hierarchical Classification.

*3.4. Hierarchical Classification*

With the training set labelled into two subsets, the process of Hierarchical Classification entails training a classification model that classifies each requirement in a hierarchical fashion, as Figure 2 shows. The main steps in hierarchical classification are:

1. At the top level, we train a *binary classifier*, $F_{super}$, to classify each requirement in the training set into either the majority class subset or the minority class subset, based on the "$maj$" and "$min$" labels, resulting in two balanced subsets.
2. At the second level, for the majority set, we train a *multiclass classifier*, $F_{maj}$, to classify each requirement into one of several categories. For the minority set, we also train a *multiclass classifier*, $F_{min}$, that respectively perform classification in its corresponding subset, to classify each requirement into one of several categories.

These three classifiers form a hierarchical classification model collectively performing multiclass classification of requirements in the training set.

We have implemented the HC4RC approach in Python programming language and made the source code of this implementation publicly available in Zenodo (Binkhonain and Zhao, 2022).

## 4. Evaluation of HC4RC

To evaluate HC4RC, we experimentally compare it with three closely related ML approaches. This involves implementing HC4RC and the three related approaches, and then comparing their performance results to assess the strengths and weaknesses of HCRC against its three peer approaches. In this section, we detail our experimental procedures and methods. We present and analyse the results obtained in Section 5.

### 4.1. Three Related ML Approaches for Evaluating HC4RC

The approaches we are looking for comparison must meet the following criteria: 1) They must be closely related; 2) They should explicitly deal with imbalanced classes or feature selection; and 3) Their description should be clear and detailed enough so that we can reimplement them or their source code is available so that we can adapt their code.

Most existing ML approaches for requirements classification, such as those included in a recent survey (Binkhonain and Zhao, 2019), do not meet our comparison criteria. Here we introduce the three selected approaches that meet our criteria.

#### 4.1.1. The K&M Approach

Proposed by Kurtanović and Maalej (2017), the K&M approach performs both binary and multiclass classification tasks. A binary classifier was trained using the PROMISE NFR dataset to classify a requirement as a FR or NFR. For multiclass classification, the K&M approach only considered the four most frequent NFRs in the PROMISE NFR dataset, i.e., Usability, Security, Operational, and Performance. Two types of classifiers were developed to perform multiclass classification: a set of binary classifiers, one for each requirements category, and one single multiclass classifier for classifying all requirements categories. Both binary and multiclass classifiers were SVM-based. However, the K&M approach only addressed the two class imbalance problem between Usability and the rest of NFRs (treating non-usability requirements as one class). Data sampling techniques were employed by adding supplementary samples derived from Amazon software reviews to the minority class (Usability) and randomly removing some samples from the majority class (non-Usability).

For feature selection, the K&M approach used different types of features, including word n-grams, POS tag based n-grams and syntactic features.

The authors of the K&M approach reported that the best performance was achieved when all the word n-gram features, with $n \in \{1, 2, 3\}$, were used for binary classification (FRs or NFRs) and the next best performance was achieved using the top 500 selected word features (out of 1,000).

*4.1.2. The NoRBERT Approach*

Proposed by Hey et al. (2020a), this transfer learning approach uses the fine-tuning technique to adopt two pre-trained **BERT** models ($BERT_{base}$ and $BERT_{large}$) (Devlin et al., 2018) for requirements classification. The PROMISE NFR dataset was also used for fine-tuning the BERT models.

Apart from using a different kind of classification model, NoRBERT had many similarities to the K&M approach: Both approaches performed both binary and multiclass classification tasks; both used the PROMISE NFR dataset as the training set; both applied under-sampling and oversampling to imbalanced classes for binary classification. However, NoRBERT outperformed the K&M approach, due to its use of a state-of-the-art deep learning model.

One issue that we found from the source code of NoRBERT (Hey et al., 2020b) was that NoRBERT used a weighted average F1 for calculating the overall performance values for both binary and multiclass classification, which is biased towards the majority classes. We will discuss this issue in Section 4.4 and propose a different way to calculate the average performance results for multiclass classifiers.

*4.1.3. The Yin Approach*

Proposed by Yin et al. (2013), the Yin approach was originally developed for binary classification of medical image data, not for requirements classification. However, we selected it to compare with our HC4RC as we were interested in its unique class decomposition technique. This decomposition technique decomposes the majority class in the dataset for a binary classification task into several relatively balanced pseudo-subclasses for the purpose of feature selection, with balanced instances in each one. Afterwards, the Yin approach applied a Hellinger distance-based feature selection technique (Cieslak et al., 2012) on the decomposed classes. This feature selection technique is said to be independent of the class distributions and can handle the high-dimensional class-imbalanced data (Fu et al., 2020). In our comparison, we reimplemented the Yin approach for multiclass classification of requirements so that it can be compared with our HC4RC.

*4.2. The Requirements Dataset*

The training set for our evaluation was the PROMISE-exp dataset (Lima et al., 2019), which is an expansion of the PROMISE NFR dataset (Cleland-Huang et al., 2007b). The original PROMISE NFR dataset contains a total of 625 labelled requirements, distributed across 12 classes, made up of one FR class and 11 NFR classes. The FR class has 255 requirements in total and the NFR classes have 370 requirements in total. These requirements are collected from 15 requirements documents (i.e., 15 software projects). This dataset has become the *de facto* dataset for training new ML approaches for requirements classification (Hussain et al., 2008; Kurtanović and Maalej, 2017; Abad et al., 2017; Dalpiaz et al., 2019).

In contrast, the PROMISE-exp dataset contains 969 requirements from 47 requirements documents. The number of classes in this dataset is the same as the original PROMISE NFR dataset. Figure 3 depicts these classes and their requirements distribution in PROMISE-exp. The FR class is denoted as Functional (F); the NFR classes are Security (SE), Usability (US), Operability (O), etc. As Figure 3 shows, the PROMISE-exp dataset is imbalanced, with the largest class, Functional (F), containing 444 requirements whereas the smallest class, Portability (PO), containing only 12 requirements. Furthermore, the dataset also exhibits the HDLSS problem as it has a sample size of 969 requirements and a feature dimension of 2133 features. Clearly, $969 << 2133$.

*4.3. Implementations*

We implemented our HC4RC approach based on the description given in the previous section. We reimplemented the K&M approach based on the source code provided by Dalpiaz et al. (2019) and applied *imbalanced-learn*[2], a python package, for random over and under-sampling of the training set. For the NoRBERT approach, we adopted the source code provided by its authors (Hey et al., 2020b). We implemented the Yin approach from scratch based on its description, as its source code is not available.

We implemented the classifiers for the HC4RC, K&M and Yin approaches using the linear SVM based on the implementation provided by scikit-learn's LinearSVC (Pedregosa et al., 2011). We fine-tuned the parameters of these classifiers using scikit-learn's GridSearchCV. For NoRBERT, we only fine-tuned the $BERT_{base}$ model due to a lack of computational resources. We trained all these approaches on the PROMISE-exp dataset. The source code of our implementations of all four approaches is available at the Code Ocean platform (`https://codeocean.com`, with `doi:10.24433/CO.6887783.v1`).

---

[2]https://pypi.org/project/imblearn/

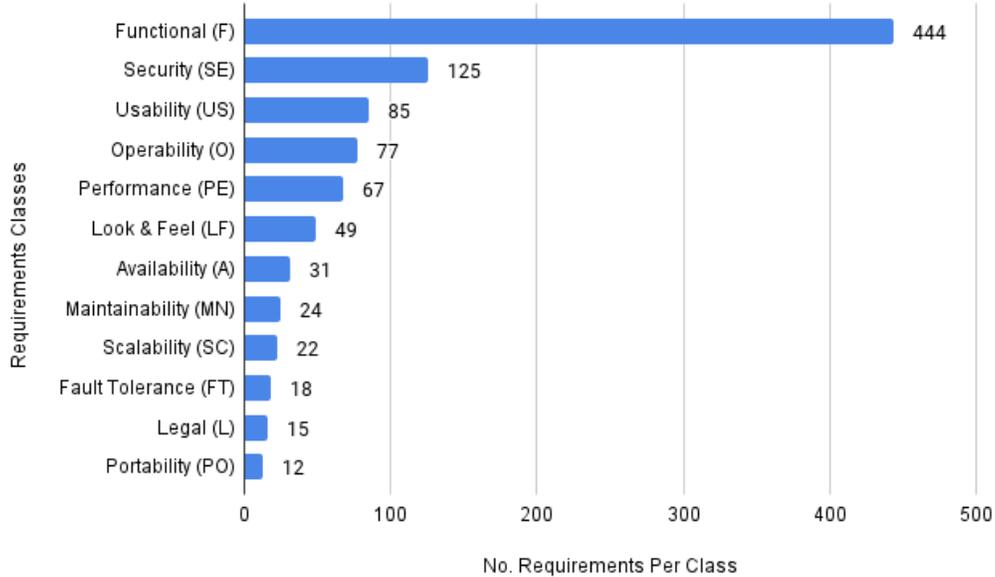

Figure 3: Requirements classes and their instances in the PROMISE-exp dataset.

We employed both *10-fold* and *p-fold* (project-specific fold) cross-validation (CV) methods to test each approach. The 10-fold CV was used to reduce the bias and variance of the approach on different parts of the data, whereas the p-fold CV was to reduce the bias and variance of the approach on different requirements projects in the dataset. In other words, *the purpose of using the 10-fold CV is to improve the generalizability of each classification approach on the unseen requirements, whereas the purpose of the p-fold CV is to improve the generalizability of each classification approach on the unseen requirements documents.* The two CV methods thus complement one another.

For 10-fold CV, we divided the PROMISE-exp dataset into 10 equal parts based on the number of requirements and executed the approach 10 times, each time using a different fold of the data for testing whereas the remaining nine parts for training. For the p-fold CV, we divided the PROMISE-exp dataset into 10 parts based on the number of projects (i.e., the number of documents in the dataset) (Cleland-Huang et al., 2007c; Dalpiaz et al., 2019). As PROMISE-exp contains 47 projects, we assigned 4−5 documents to each fold. The p-fold CV process is the same as the 10-fold CV. We used scikit-learn's StratifiedKFold to divide the dataset into 10-fold and p-fold.

Table 2: Computation efficiency of HC4RC, K&M, Yin, and NoRBERT

|  | HC4RC | K&M | Yin | NoRBERT |
|---|---|---|---|---|
| Execution Time | 10.73 seconds | 23.70 minutes | 10.52 seconds | 1.09 hours |
| Memory load | 1.7 GB | 3.9 GB | 0.75 GB | 5.7 GB |

We carried out the training and testing of all four approaches on a standard laptop with an Intel Core i5 1.6 GHz and 8 GB RAM. The computation efficiency of each approach was measured by the time taken and the memory used to train and test the approach. The Python *time* and *psutil* libraries were respectively used for measuring the execution time and memory usage. The measurements for these four approaches are given in Table 2. The table shows that for the execution time, the Yin approach is the fasted one, followed by HC4RC and then K&M; NoRBERT takes the longest time. For memory load, Yin consumes the least memory space whereas NoRBERT consumes the most.

We measured the classification performance (effectiveness) of each approach using the metrics described in the section below.

### *4.4. Evaluation Metrics*

We measure the classification performance of each approach on individual classes using the unweighted precision ($P$), recall ($R$) and F-1 score ($F1$). These metrics calculate the performance of an approach on each individual class by statistically comparing the predicted class for each requirement with its true label.

We then measure the overall performance of each approach on all the classes using the recommended multiclass classification metrics of *macro* and *micro average P, R* and *F1* (Grandini et al., 2020). Macro Average P and R are simply computed as the arithmetic mean of the metrics for individual classes:

$$\textit{Macro-Average-P} = \frac{\sum_{k=1}^{K} P_k}{K}, \tag{1}$$

$$\textit{Macro-Average-R} = \frac{\sum_{k=1}^{K} R_k}{K} \tag{2}$$

Macro F1 is the harmonic mean of Macro-Precision and Macro-Recall:

$$Macro\text{-}F1 = 2 * (\frac{MacroAverage\text{-}P * Macro\text{-}Average\text{-}R}{Macro\text{-}Average\text{-}P^{-1} + Macro\text{-}Average\text{-}R^{-1}}) \qquad (3)$$

Micro-Average P, R and F1 are equal to Accuracy as follows:

$$Micro\text{-}Average\text{-}P = Micro\text{-}Average\text{-}R = Micro\text{-}Average\text{-}F1 = \frac{\sum_{k=1}^{K} TP_k}{\text{Grand Total}} \qquad (4)$$

In the above formulae, $K$ is the number of classes in the dataset, whereas $k$ denotes an individual class. These formulae are explained as follows:

*Macro average P, R and F1* evaluate the performance of a multiclass approach at the class level, without consideration of the size of classes. Under these metrics, a higher macro-F1 score indicates that the approach performs well on all the classes, regardless of large or small, whereas a lower macro-F1 score indicates the poorer performance of the approach (Grandini et al., 2020).

On the other hand, *micro average P, R, and F1* are all measured using the same Accuracy metric and thus have the same score (Grandini et al., 2020). Furthermore, these metrics evaluate a multiclass approach by considering the size of each class and thus they give more importance to majority classes. In other words, for micro averages, poor performance on small classes is not so important, as the number of instances belonging to those classes is small compared to the overall number of instances in the dataset (Grandini et al., 2020). Under these metrics, a higher micro-F1 score indicates that the approach is more accurate overall, whereas a lower macro-F1 score indicates the approach is less accurate overall. Thus *macro average F1* and *micro average F1* complement one another in that the former measures each class equally, whereas the latter measures each instance equally (Sokolova and Lapalme, 2009).

Based on the aforementioned evaluation metrics, we measure the classification performance of the HC4RC, K&M, Yin, and NoRBERT approaches. The measurements are presented in Table 3 while Figure 4 depicts the macro and micro averages of these approaches. We discuss these results in the next section.

## 5. Results Analysis and Discussion

The classification performance results obtained by the four approaches are presented in Table 3. In this section, we compare, analyze and interpret these results. Where appropriate, we explain why our approach performs better or worse than its peer approaches.

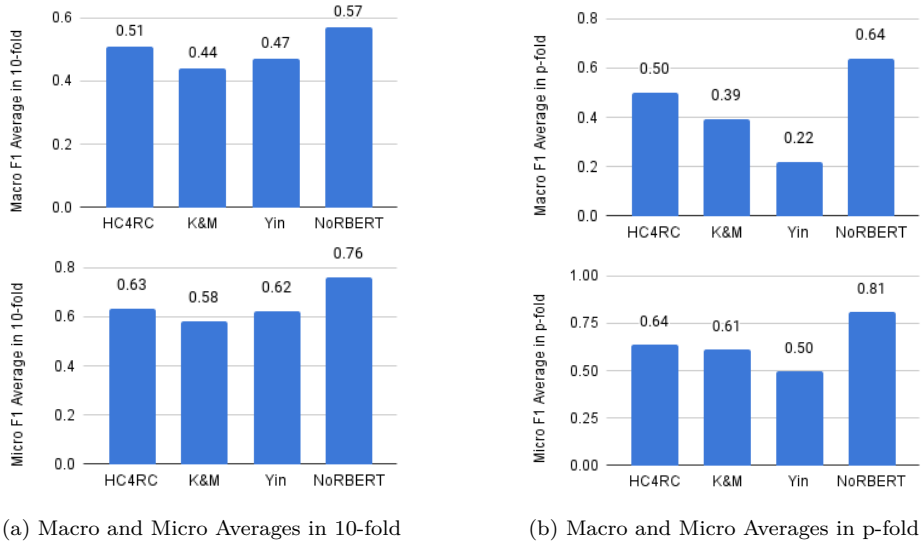

(a) Macro and Micro Averages in 10-fold

(b) Macro and Micro Averages in p-fold

Figure 4: Macro and micro averages of the four approaches on classification of 12 requirements classes.

## 5.1. Comparing HC4RC with K&M

Table 3 shows that HC4RC outperformed K&M on all but one class in both 10-fold and p-fold CV. In both cases the class under-performed is a minority class, i.e., Portability (PO) in 10-fold whereas Legal (L) in p-fold. As both HC4RC and K&M are based on the SVM model, the better performance achieved by our approach suggests that our hierarchical classification approach incorporating semantic role-based feature selection and dataset decomposition can handle imbalanced classes better than the K&M approach.

Figure 4 shows that HC4RC has higher macro and micro averages than K&M. In particular, macro averages show that HC4RC outperformed K&M considerably on individual classes (0.51 vs 0.44 in 10-fold and 0.50 vs 0.39 in p-fold), whereas micro averages show that HC4RC achieved an overall better performance than K&M on all 12 classes (0.63 vs 0.58 in 10-fold and 0.64 vs 0.61 in p-fold). These results also show that HC4RC has a better generalizability on both unseen requirements (10-fold) as well as unseen requirements projects (p-fold) than K&M.

Table 3: Classification performance of HC4RC, K&M, Yin, and NoRBERT on 12 requirements classes in the PROMISE-exp dataset. The highest F1 score for each class is in bold.

| | HC4RC | | | K&M | | | Yin | | | NoRBERT | | |
|---|---|---|---|---|---|---|---|---|---|---|---|---|
| *Class* | *P* | *R* | *F1* | *P* | *R* | *F1* | *P* | *R* | *F1* | *P* | *R* | *F1* |
| **10-fold cross validation** | | | | | | | | | | | | |
| F (444) | 0.76 | 0.73 | 0.74 | 0.69 | 0.76 | 0.73 | 0.69 | 0.82 | 0.75 | 0.88 | 0.90 | **0.89** |
| SE (125) | 0.69 | 0.60 | 0.64 | 0.62 | 0.55 | 0.58 | 0.61 | 0.57 | 0.59 | 0.78 | 0.87 | **0.83** |
| US (85) | 0.44 | 0.58 | 0.50 | 0.51 | 0.35 | 0.42 | 0.56 | 0.35 | 0.43 | 0.61 | 0.66 | **0.63** |
| O (77) | 0.37 | 0.66 | 0.47 | 0.36 | 0.32 | 0.34 | 0.42 | 0.42 | 0.42 | 0.53 | 0.58 | **0.56** |
| PE (67) | 0.37 | 0.66 | 0.47 | 0.52 | 0.58 | 0.55 | 0.69 | 0.55 | 0.61 | 0.73 | 0.79 | **0.76** |
| LF (49) | 0.42 | 0.53 | 0.47 | 0.36 | 0.33 | 0.34 | 0.33 | 0.31 | 0.32 | 0.53 | 0.53 | **0.53** |
| A (51) | 0.84 | 0.68 | **0.75** | 0.53 | 0.58 | 0.55 | 0.45 | 0.61 | 0.52 | 0.68 | 0.55 | 0.61 |
| MN (24) | 0.70 | 0.29 | **0.41** | 0.23 | 0.25 | 0.24 | 0.67 | 0.25 | 0.36 | 0.29 | 0.25 | 0.27 |
| SC (22) | 0.45 | 0.41 | 0.43 | 0.53 | 0.45 | 0.49 | 0.79 | 0.68 | **0.73** | 0.80 | 0.36 | 0.50 |
| FT (18) | 0.83 | 0.28 | 0.42 | 0.29 | 0.28 | 0.29 | 0.83 | 0.28 | 0.42 | 0.80 | 0.53 | **0.64** |
| L (15) | 0.64 | 0.47 | **0.54** | 0.60 | 0.40 | 0.48 | 0.50 | 0.08 | 0.14 | 0.75 | 0.40 | 0.52 |
| PO (12) | 0.50 | 0.08 | 0.14 | 0.25 | 0.25 | 0.25 | 0.33 | 0.17 | 0.22 | 0.75 | 0.17 | **0.27** |
| *Macro* | 0.61 | 0.48 | 0.51 | 0.45 | 0.43 | 0.44 | 0.56 | 0.43 | 0.47 | 0.66 | 0.55 | **0.57** |
| *Micro* | 0.63 | 0.63 | 0.63 | 0.58 | 0.58 | 0.58 | 0.62 | 0.62 | 0.62 | 0.76 | 0.76 | **0.76** |
| **p-fold cross validation** | | | | | | | | | | | | |
| F (444) | 0.79 | 0.73 | 0.76 | 0.68 | 0.86 | 0.76 | 0.55 | 0.86 | 0.67 | 0.90 | 0.92 | **0.91** |
| SE (125) | 0.64 | 0.65 | 0.65 | 0.64 | 0.60 | 0.62 | 0.43 | 0.26 | 0.32 | 0.82 | 0.88 | **0.85** |
| US (85) | 0.40 | 0.62 | 0.49 | 0.49 | 0.33 | 0.39 | 0.44 | 0.19 | 0.26 | 0.68 | 0.74 | **0.71** |
| O (77) | 0.32 | 0.61 | 0.42 | 0.38 | 0.33 | 0.35 | 0.22 | 0.14 | 0.17 | 0.66 | 0.76 | **0.71** |
| PE (67) | 0.64 | 0.55 | 0.59 | 0.70 | 0.45 | 0.55 | 0.34 | 0.14 | 0.19 | 0.81 | 0.85 | **0.83** |
| LF (49) | 0.46 | 0.53 | 0.50 | 0.36 | 0.33 | 0.34 | 0.33 | 0.31 | 0.32 | 0.53 | 0.53 | **0.53** |
| A (51) | 0.76 | 0.58 | 0.66 | 0.59 | 0.53 | 0.56 | 0.27 | 0.09 | 0.14 | 0.65 | 0.61 | **0.63** |
| MN (24) | 0.55 | 0.23 | 0.32 | 0.21 | 0.20 | 0.20 | 0.21 | 0.12 | 0.15 | 0.59 | 0.38 | **0.47** |
| SC (22) | 0.56 | 0.38 | 0.45 | 0.55 | 0.35 | 0.42 | 0.14 | 0.04 | 0.06 | 0.60 | 0.38 | **0.46** |
| FT (18) | 0.75 | 0.19 | 0.30 | 0.29 | 0.17 | 0.22 | 0.43 | 0.19 | 0.26 | 0.92 | 0.73 | **0.81** |
| L (15) | 1.00 | 0.07 | 0.13 | 0.25 | 0.06 | 0.10 | 0.00 | 0.00 | 0.00 | 0.60 | 0.25 | **0.35** |
| PO (12) | 0.67 | 0.17 | 0.27 | 0.27 | 0.21 | 0.23 | 0.33 | 0.06 | 0.11 | 0.55 | 0.33 | **0.41** |
| *Macro* | 0.61 | 0.48 | 0.50 | 0.45 | 0.35 | 0.39 | 0.31 | 0.19 | 0.22 | 0.70 | 0.61 | **0.64** |
| *Micro* | 0.64 | 0.64 | 0.64 | 0.61 | 0.61 | 0.61 | 0.50 | 0.50 | 0.50 | 0.81 | 0.81 | **0.81** |

*5.2. Comparing HC4RC with Yin*

Table 3 shows that HC4RC also outperformed the Yin approach on all but one class in 10-fold and it outperforms the Yin approach on all classes in p-fold CV. Both approaches applied decomposition to address the class imbalance problem. However, the better performance of our approach indicates that its data decomposition technique works better in handling imbalance problems in multi-classification problems than the Yin approach's class decomposition.

Figure 4 shows that HC4RC has higher macro and micro averages than Yin. In particular, macro averages show that HC4RC outperformed Yin considerably on individual classes (0.51 vs 0.47 in 10-fold and 0.50 vs 0.22 in p-fold), whereas micro averages show that HC4RC achieved an overall better performance than Yin on all 12 classes (0.63 vs 0.62 in 10-fold and 0.64 vs 0.50 in p-fold).

These results also show that HC4RC has a better generalizability on both unseen requirements (10-fold) as well as unseen requirements projects (p-fold) than Yin.

*5.3. Comparing HC4RC with NoRBERT*

Table 3 shows that NoRBERT outperformed HC4RC on almost all classes under both 10-fold and p-fold CV. However, we notice that NoRBERT's performance fluctuates in small classes. In particular, NoRBERT performed worse than HC4RC on the Maintainability (MN) class under 10-fold and on the Availability (A) under p-fold. The poorer performance of NoRBERT on some minority classes indicates that even the state-of-the-art deep learning model is still limited when it comes to classifying small classes in multiclass classification.

Figure 4 shows that HC4RC performed worse than NoRBERT on individual classes as well as on all 12 classes as a whole. These results clearly show that the combination of class imbalance and HDLSS data has seriously degraded the classification performance of classical statistical methods such as the SVM model used in HC4RC.

These results also show that NoRBERT has a much better generalizability on both unseen requirements (10-fold) as well as unseen requirements projects (p-fold) than HC4RC.

*5.4. Further Analysis and Discussion*

We first compare the overall classification performance of these four approaches from the viewpoint of their macro and micro average scores (see

Figure 4). The findings are discussed as follows.

Of the four compared approaches, NoRBERT is the best overall approach for requirements classification. Its macro averages show that NoRBERT has achieved the best performance on individual classes and its micro averages show that it has achieved the best performance on all 12 classes. Furthermore, its performance results in 10-fold and p-fold show that NoRBERT has the best generalizability on both unseen requirements and unseen requirements projects. These results also suggest that NoRBERT is the best approach for dealing with class imbalance and HDLSS data in requirements documents. As NoRBERT applied the same data sampling techniques to imbalanced data as K&M, we assume that *the strong performance of NoRBERT is due to its underlying deep learning model BERT*.

Of the three approaches that used the SVM model as their classification model (HC4RC, K&M and Yin), HC4RC is the best overall approach for requirements classification. Its macro averages show that HC4RC has achieved the best performance on individual classes and its micro averages show that it has achieved the best performance on all 12 classes. Furthermore, its performance results in 10-fold and p-fold show that HC4RC has the best generalizability on both unseen requirements and unseen requirements projects. These results also suggest that HC4RC is the best SVM-based approach for dealing with class imbalance and HDLSS data in requirements documents. As the three SVM-based approaches applied different techniques for handling class imbalance and HDLSS data, we assume that *our semantic role-based feature selection combined with dataset decomposition and hierarchical classification is more effective than the data sampling and hybrid feature selection techniques used in K&M, and class decomposition and Hellinger distance-based feature selection in the Yin approach.*

We now look into the performance of these four approaches on individual classes from the viewpoint of their unweighted P, R and F1 scores (see Table 3). The findings are discussed as follows.

First, on the largest class F (functional requirements), all four approaches performed relatively well. NoRBERT in particular achieved a F1 score of 0.89 in 10-fold and 0.91 in 10-fold. Both HC4RC and K&M achieved a F1 score of 0.74 and 0.73 respectively in 10-fold and 0.76 in p-fold. Yin achieved 0.73 in 10-fold and 0.67 in p-fold. These scores also show that NoRBERT, HC4RC and K&M have a better generalizability on the unseen requirements projects than the unseen requirements. We believe this ability is critically important for requirements classification, as a ML approach should be able to differentiate requirements in different projects.

Second, on the remaining classes (SE, US, etc.), NoRBERT performed better in p-fold than 10-fold; HC4RC and K&M performed similarly in 10-fold and p-fold; Yin performed better in 10-fold than p-fold. These results suggest that NoRBERT has better generalizability on the unseen requirements projects than the unseen requirements; HC4RC and K&M have similar generalizability on the unseen requirements and the unseen requirements documents; Yin has a better generalizability on the unseen requirements than the unseen requirements documents.

Under 10-fold CV, we notice that HC4RC outperformed the rest three approaches on three small classes A, MN and L; Yin outperformed the rest three approaches on one small class SC; NoRBERT outperformed the rest three approaches on two small classes FT and PO. These results show that *HC4RC achieved a better performance than NoRBERT on the small classes for the unseen requirements*.

Under p-fold CV, we notice that HC4RC outperformed the rest three approaches on every class, but it suffered the performance loss on the small classes MN, SC, L, and PO. These results show that *the classification performance of the deep learning model BERT has also been degraded on HDLSS requirements*, a key finding from our evaluation.

*Finally, we attribute the better performance of HC4RC than K&M and Yin to the aggregate effect of the three key techniques employed by HC4RC.*

## 6. Threats to Validity

In this section, we discuss potential threats to the validity of our evaluation of HC4RC and explain why we believe such threats are minimal.

*Reimplementation of related approaches.* One potential validity threat is our reimplementations of the K&M, Yin and NoRBERT approaches, as we modified these approaches so that they can be used to perform multiclass classification on the same dataset. While we cannot avoid this threat entirely, we are making the source code for our reimplementations of these approaches publicly available (Binkhonain and Zhao, 2022), so that other researchers can assess its validity.

*The quality of the training set.* The PROMISE NFR dataset on which the PROMISE-exp dataset was built was known for its mislabelling issues, as the dataset was labelled by students (Hey et al., 2020a). However, while we believe that the poor quality of the dataset can affect the performance of the approaches in our evaluation, it should not affect the generalizability of these approaches, as we applied this dataset consistently to all the approaches. Furthermore, since both the original PROMISE NFR and PROMISE-exp datasets have been widely used in the RE community, using them for research evaluation should be a strength, not a weakness, as they allow us to compare our results directly to other approaches (Kurtanović and Maalej, 2017; Hey et al., 2020a). We make our code and data publically available so that further replication or reproduction of our approach can be carried out. We recognize that the lack of the gold standard labelled requirements datasets has been an open challenge to using ML approaches for RE tasks (Binkhonain and Zhao, 2019).

*Performance measure.* Another concern is how well metrics can really measure what they are intended to measure (Ralph and Tempero, 2018). In RE, we noticed that researchers normally use unweighted F1 (Kurtanović and Maalej, 2018; Dalpiaz et al., 2019) or weighted *F1* (Hey et al., 2020a) to measure the performance of both binary and multiclass classifiers. For example, Hey et al. (2020a) used the weighted average *F1*-score over all classes weighted by the frequency of appearance - that is, to weigh larger classes more than smaller classes. We believe such a weighting can inflate the performance of the larger classes and skew the overall results. In our evaluation, we rationally chose macro and micro average metrics to evaluate and compare different multiclass approaches, to avoid the bias towards large classes (Grandini et al., 2020), as discussed in Section 4. However, we agree that in RE, achieving higher recall is more important than higher precision and in this context, a weighted *F1*-score that gives more importance to $R$ than $P$ is desirable (Berry, 2021). While we have not used such a weighted metric in our experiments, we believe our metrics have minimized the classifier bias.

*Generalizibility.* As our comparison has been limited to a single dataset, a potential threat is the validity of our evaluation conclusion. To mitigate this threat, we used both 10-fold and *p*-fold CV to test all the approaches, as these cross-validation methods were designed for evaluating ML models on limited data samples (Bengio et al., 2003).

## 7. Conclusion

This paper proposes HC4RC, a novel machine learning approach for multiclass classification of requirements. HC4RC is designed to address two specific problems in requirements classification: class imbalance and HDLSS. These problems, common to requirements classification tasks, can greatly degrade the performance of ML methods. HC4RC solves the first problem through dataset decomposition and hierarchical classification; it deals with the second problem through a novel semantic role-based feature selection method. The novelty of HC4RC thus lies in its combination of these three techniques into a simple and practical approach that can effectively solve the problems of class imbalance and HDLSS. The key findings of this paper are summarized as follows:

- Overall, HC4RC performs better than the two SVM-based approaches, K&M and Yin, and performs only slightly worse than the BERT-based approach, NoRBERT. This finding shows that our semantic role-based feature selection combined with dataset decomposition and hierarchical classification provides a more effective solution to class imbalance and HDLSS data than the data sampling and hybrid feature selection techniques used in K&M, and class decomposition and Hellinger distance-based feature selection in the Yin approach. As NoRBERT applied the same data sampling techniques to imbalanced data as K&M, we assume that the strong performance of NoRBERT is due to its underlying deep learning model BERT.

- On individual classes, HC4RC performs better than all other compared approaches on small classes for the unseen requirements. This shows that the classification performance of the deep learning based NoRBERT can also be degraded on imbalanced classes.

- HC4RC has a better generalizability on both unseen requirements (shown in 10-fold CV) as well as unseen requirements projects (shown in p-fold CV) than its closest peers, K&M and Yin, but it is worse than NoRBERT. This means that the DL based approach has a better generalizability than the traditional ML approach.

In conclusion, our results, while still preliminary as they are based only on one dataset, suggest that multiclass classification of requirements with the class imbalance and HDLSS problems presents a challenge to ML approaches in general, even for the advanced deep learning models. This paper has made a *practical* contribution to addressing these problems. We suggest future work on requirements classification to focus on the following areas:

- For ML approaches based on the traditional statistical classification models, more work is needed to develop better feature selection techniques such as semantic representations and roles of requirements. We believe our work presented in this paper has made a start in this area.

- For ML approaches based on the advanced deep learning models, more work is needed to train these models on requirements specific data. Some pioneering work has already started in this area (Ajagbe and Zhao, 2022).

- For both traditional and advanced learning approaches, more work is needed to investigate different data re-balancing techniques, such as those presented in this paper.

**Acknowledgments**

We wish to thank the three reviewers for their expert comments to our paper. We thank the University of Manchester for providing an Open Access fund for this paper.